\documentclass[twocolumn,trackchanges]{aastex63}

\expandafter\def\csname editcolor1\endcsname{magenta}% default is black
\expandafter\def\csname editcolor2\endcsname{violet}% default is black

\usepackage{amsmath}
\usepackage{xspace}
\usepackage{multirow}
\usepackage{fancyapj}
\usepackage[modulo]{lineno}

\newcommand*\patchAmsMathEnvironmentForLineno[1]{%
  \expandafter\let\csname old#1\expandafter\endcsname\csname #1\endcsname
  \expandafter\let\csname oldend#1\expandafter\endcsname\csname end#1\endcsname
  \renewenvironment{#1}%
     {\linenomath\csname old#1\endcsname}%
     {\csname oldend#1\endcsname\endlinenomath}}% 
\newcommand*\patchBothAmsMathEnvironmentsForLineno[1]{%
  \patchAmsMathEnvironmentForLineno{#1}%
  \patchAmsMathEnvironmentForLineno{#1*}}%
\AtBeginDocument{%
\patchBothAmsMathEnvironmentsForLineno{equation}%
\patchBothAmsMathEnvironmentsForLineno{align}%
\patchBothAmsMathEnvironmentsForLineno{flalign}%
\patchBothAmsMathEnvironmentsForLineno{alignat}%
\patchBothAmsMathEnvironmentsForLineno{gather}%
\patchBothAmsMathEnvironmentsForLineno{multline}%
}

\renewenvironment{quote}%
  {\list{}{\leftmargin=0.1in\rightmargin=0.1in}\item[]}%
  {\endlist}

\newcommand{\E}[1]{\ensuremath{\times 10^{#1}} }

\newcommand{\msol}{\ensuremath{M_{\odot}}\xspace}

\newcommand{\per}[1]{\,#1\ensuremath{^{-1}}\xspace}
\newcommand{\persq}[1]{\,#1\ensuremath{^{-2}}\xspace}
\newcommand{\cts}{ct\per{s}}
\newcommand{\lumcgs}{{\rm erg\per{s}}\xspace}
\newcommand{\fluxcgs}{{\rm erg\per{s}\persq{cm}}\xspace}

\newcommand{\s}{{\rm\,s}\xspace}
\newcommand{\kev}{{\rm\,keV}\xspace}
\newcommand{\phnorm}{{\rm\,ph\persq{cm}\per{s}}\xspace}

\newcommand{\rxte}{\textit{RXTE}\xspace}

\newcommand{\nicer}{\textit{NICER}\xspace}

\newcommand{\src}{SAX~J1808\xspace}

\newcommand{\rchi}{\ensuremath{\chi^2_r}\xspace}

\begin{document}

\title{A NICER thermonuclear burst from the millisecond X-ray pulsar SAX J1808.4--3658}

\author{Peter Bult}
\affiliation{Astrophysics Science Division, 
  NASA's Goddard Space Flight Center, Greenbelt, MD 20771, USA}

\author{Gaurava K. Jaisawal}
\affil{National Space Institute, Technical University of Denmark, 
  Elektrovej 327-328, DK-2800 Lyngby, Denmark}

\author{Tolga G{\"u}ver}
\affiliation{Department of Astronomy and Space Sciences, Science Faculty, 
  Istanbul University, Beyaz{\i}t, 34119 Istanbul, Turkey}
\affiliation{Istanbul University Observatory Research and Application Center, 
  Beyaz{\i}t, 34119 Istanbul, Turkey}

\author{Tod E. Strohmayer} 
\affil{Astrophysics Science Division and Joint Space-Science Institute,
  NASA's Goddard Space Flight Center, Greenbelt, MD 20771, USA}

\author{Diego Altamirano}
\affiliation{Physics \& Astronomy, University of Southampton, 
  Southampton, Hampshire SO17 1BJ, UK}

\author{Zaven Arzoumanian} 
\affiliation{Astrophysics Science Division, 
  NASA's Goddard Space Flight Center, Greenbelt, MD 20771, USA}

\author{David R. Ballantyne}
\affil{Center for Relativistic Astrophysics, School of Physics, 
  Georgia Institute of Technology, 837 State Street, Atlanta, GA 30332-0430, USA}

\author{Deepto Chakrabarty}
\affil{MIT Kavli Institute for Astrophysics and Space Research, 
  Massachusetts Institute of Technology, Cambridge, MA 02139, USA}

\author{J\'er\^ome Chenevez}
\affil{National Space Institute, Technical University of Denmark, 
  Elektrovej 327-328, DK-2800 Lyngby, Denmark}

\author{Keith C. Gendreau} 
\affiliation{Astrophysics Science Division, 
  NASA's Goddard Space Flight Center, Greenbelt, MD 20771, USA}

\author{Sebastien Guillot} 
\affil{CNRS, IRAP, 9 avenue du Colonel Roche, BP
  44346, F-31028 Toulouse Cedex 4, France} 
\affil{Universit\'e de Toulouse, CNES, UPS-OMP, F-31028 Toulouse, France}

\author{Renee M. Ludlam}
\altaffiliation{Einstein Fellow}
\affiliation{Cahill Center for Astronomy and Astrophysics, 
  California Institute of Technology, Pasadena, CA 91125, USA}

\begin{abstract}
    The \textit{Neutron Star Interior Composition Explorer} (\nicer) has
    extensively monitored the August 2019 outburst of the 401 Hz 
    millisecond X-ray pulsar SAX J1808.4--3658. In this Letter, we report on
    the detection of a bright helium-fueled Type~I X-ray burst.
    With a bolometric peak flux of $(2.3\pm0.1)\E{-7}$\,\fluxcgs, this
    was the brightest X-ray burst among all bursting sources observed with
    \nicer to date.  The burst shows a remarkable two-stage evolution in flux,
    emission lines at $1.0$ keV and $6.7$ keV, and burst
    oscillations at the known pulsar spin frequency, with $\approx4$\%
    fractional sinusoidal amplitude.  We interpret the burst flux evolution as the
    detection of the local Eddington limits associated with the hydrogen and
    helium layers of the neutron star envelope. The emission lines are likely
    associated with Fe, due to reprocessing of the burst emission
    in the accretion disk.
\end{abstract}

\keywords{
stars: neutron ---
X-rays: binaries ---	
X-rays: individual (SAX J1808.4--3658)
}

\section{Introduction}
  \label{sec:intro}
  The proto-typical accreting millisecond X-ray pulsar (AMXP) SAX
  J1808.4--3658 (hereafter \src), was first discovered through the
  detection of a thermonuclear (Type~I) X-ray burst with the
  \textit{BeppoSAX} satellite in September 1996 \citep{Zand1998}. With
  X-ray outbursts recurring every $2-4$ years, this source has been
  extensively monitored ever since, leading to the first detection of
  accretion-powered millisecond pulsations \citep{Wijnands1998a}, and the
  confirmation that X-ray burst oscillations correspond with the stellar
  spin frequency \citep{Chakrabarty2003}. 
  
  In each of the eight outbursts from \src that occurred between 1996 and 2015,
  at least one X-ray burst has been detected \citep{Zand2001a, Bult2015b,
  Patruno2017b, Sanna2017c}. The majority of these bursts showed burst oscillations
  \citep{Bilous2019} and were observed near peak luminosity of their respective
  outbursts, when the accretion rate was $\approx(3-5)\E{-10}$\,\msol\per{yr}
  \citep{Bult2015b}. Detailed modelling of a well sampled burst train observed
  with the \textit{Rossi X-ray Timing Explorer} (\rxte) in October 2002
  \citep{Galloway2006} demonstrated that these events are examples of X-ray
  bursts in the ``delayed helium" regime \citep{Narayan2003,Galloway2017}. In
  brief, these bursts are due to a thermonuclear flash in a nearly pure helium
  layer of the neutron star envelope. This layer of helium builds up on a
  timescale of one to a few days, through a stable $\beta$-limited CNO cycle at
  the base of a hydrogen layer. The hydrogen layer, in turn, is replenished by
  the continuous accretion of gas supplied by a hydrogen-rich brown dwarf
  companion star, which resides in a 2.01 hour orbit around the pulsar
  \citep{Chakrabarty1998, Bildsten2001}.

  The X-ray bursts of \src have also reliably shown photospheric radius
  expansion (PRE; see, e.g., \citealt{Galloway2008}, for a review). Such PRE may
  drive the ejection of burning ashes, whose presence could cause discrete
  spectral features in the burst emission \citep{Weinberg2006, Yu2018}.
  Measuring such spectral lines gives a window into the thermonuclear burning
  reactions, and can potentially be used to constrain the neutron star
  compactness. Additionally, a large fraction of the burst emission is expected
  to be intercepted and reprocessed by the accretion disk
  \citep{Ballantyne2005, Degenaar2018, Fragile2018}, providing an opportunity to
  characterize the state of the accretion disk through the spectrum of the
  reflected burst emission.

  Launched in June 2017, the \textit{Neutron Star Interior Composition
  Explorer} (\nicer; \citealt{Gendreau2017}) combines good spectral
  resolution with superb time-resolution and high throughput in the
  ${0.2-12}$\kev energy band. These properties make \nicer an ideal
  instrument to study the evolution of PRE in Type~I X-ray bursts \citep[see,
  e.g.,][]{Keek2018b, Keek2018a, Jaisawal2019}, and search for discrete
  spectral features \citep{Strohmayer2019}. 
  Hence, when \src began a new outburst in August 2019 
  \citep{ATelRussell19, ATelGoodwin19, ATelParikh19c, ATelBult19c},
  we triggered an extensive \nicer monitoring campaign. During this
  campaign we detected two X-ray bursts; the first occurred on August 9 and 
  was relatively faint, the second was seen on August 21 and was much
  brighter. In this Letter we report on the unusual properties
  of the August 21 X-ray burst. The detailed analysis of the earlier
  burst and the full \nicer campaign will be presented elsewhere. 

\section{Observations}
\label{sec:observations}
  On 2019 August 21 at 02:04 UTC, \nicer observed a bright X-ray burst
  from \src. These data are available under the \nicer ObsID 2584010501.
  We processed the data using \textsc{nicerdas} v6a, which is
  packaged with \textsc{heasoft} v6.26. We applied standard screening
  criteria, keeping only those time intervals when the pointing offset was
  $<54\arcsec$, the Earth limb elevation angle was $>15\arcdeg$, the
  elevation angle with respect to the bright Earth limb was
  $>30\arcdeg$, and the instrument was outside of the geographic region
  of the South Atlantic Anomaly. Additionally, standard background
  screening criteria were applied, which reject all epochs where the rate
  of saturating particle events (overshoots) is greater than $1$\,\cts\per{detector},
  or greater than 1.52 * \textsc{cor\_sax}$^{-0.633}$, where
  \textsc{cor\_sax}\footnote{%
      The \textsc{cor\_sax} parameter is based on a model for the cut-off
      rigidity that was originally developed for the \textit{BeppoSAX}
      satellite, and has no specific relation to \src.
  } gives the cut-off rigidity of the Earth's
  magnetic field, in units of GeV\per{c}. 
  We then applied the \textsc{barycorr} tool to correct the observed event
  times to the Solar System barycenter, where we used the JPL DE405 planetary
  ephemeris \citep{Standish1998} and the optical coordinates of
  \citet{Hartman2008}.  Finally, we estimated the background contributions to
  our data from \nicer observations of the \rxte blank-field regions
  \citep{Jahoda2006}.

\section{Analysis and Results}
\label{sec:results}
  The X-ray burst onset, $t_0$, occurred on MJD 58716.089362 TDB,
  which was $442\s$ into a $1063\s$ continuous exposure. In the following we
  focus our analysis on this exposure and express all times with respect to the
  noted onset time.

\subsection{Light curve and phenomenology}
  At $t_0$ the $0.3-10$ keV count-rate increased rapidly from an averaged
  $125$ \cts to $\approx34,000$\,\cts over a timespan of $\approx4.3\s$. The peak rate
  was maintained for $\approx3.6\s$, before the burst began to decay. The subsequent
  decay progressed on a minute-long time scale: at $t\simeq64\s$ the burst rate
  had dropped to below 5\% of the peak rate, and by the end of the available exposure,
  at $t=621\s$, the source flux had fallen to $172$ \cts. 
  {
    While this rate was slightly higher than the averaged preburst rate, the 
    preburst light curve showed a modest upward trend. If we fit this trend
    with a linear function, then we find that the burst rate decayed 
    to the extrapolated intensity at $t\simeq580\s$.
  }

\begin{figure}[t]
      \centering
      \includegraphics[width=\linewidth]{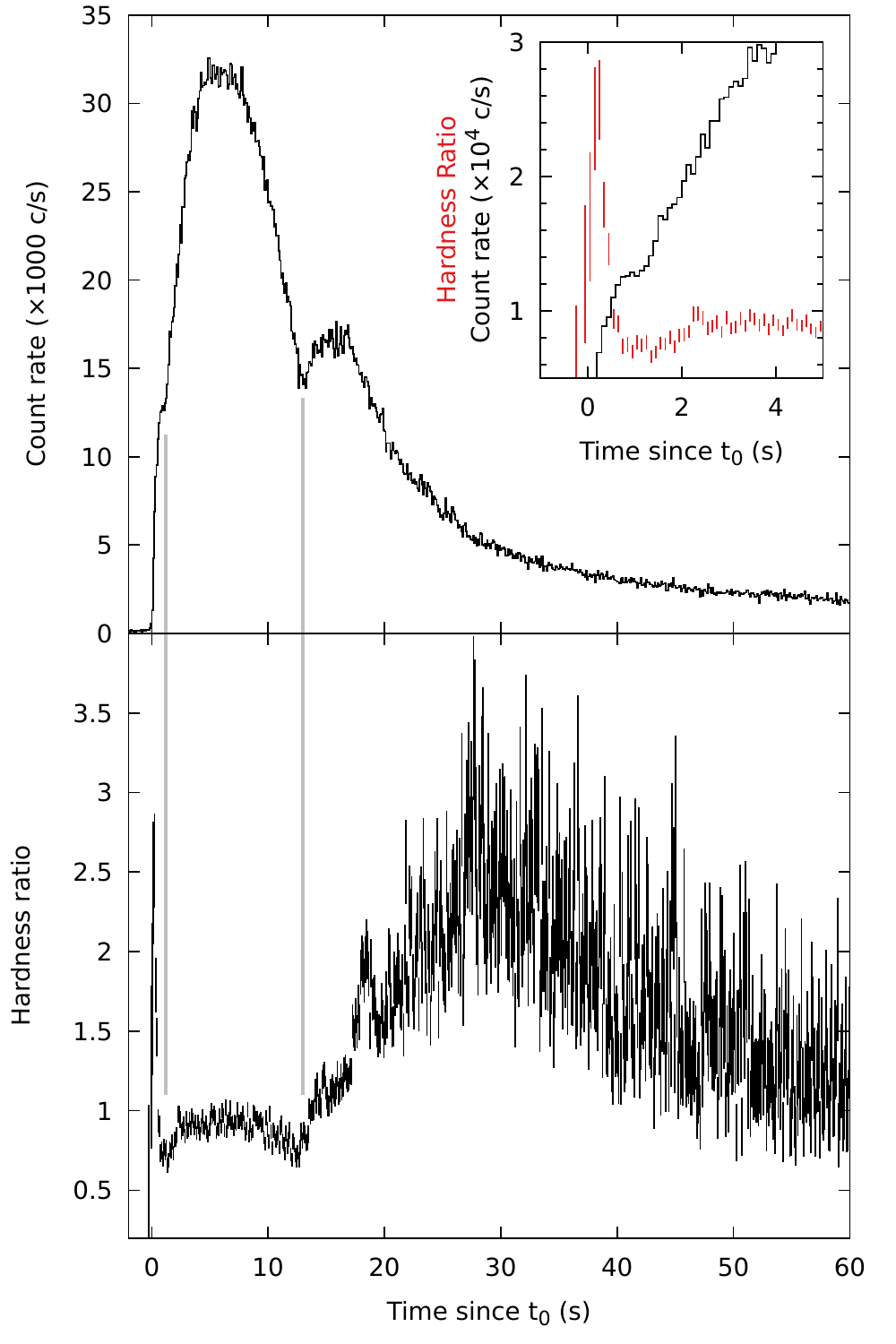}
      \caption{%
          Top: light curve of the X-ray burst from \src
          at $0.1\s$ time-resolution in the $0.3-10$ keV energy band. 
          Bottom: hardness ratio, defined as the $3-10$ keV rate divided by the
          $0.3-1$ keV rate. 
          Inset: first four seconds of the same data, with the light curve in
          black (connected line, units of $\times10^{4}$\,\cts) and the hardness ratio
          in red (vertical dashes). All panels are relative to $t_0=58716.089362$\,TDB.
          Vertical gray lines were added to guide the eye.
      }
      \label{fig:lc}
  \end{figure}

  Two unusual features stand out in the burst light curve (Figure \ref{fig:lc}). 
  First, it shows a pronounced double peaked structure, with 
  a local minimum of $\approx15,000$\,\cts at $t\simeq13.1\s$ and a secondary peak 
  of $\approx16,500$\,\cts at $t\simeq15.5\s$.
  While double-peaked X-ray bursts have commonly been
  observed, these structures are usually caused by PRE: the temperature of the
  photosphere temporarily shifts out of the instrument passband, causing an
  apparent dip in the observed X-ray rate \citep{Grindlay1980}.  Given its
  low-energy coverage, \nicer is able to follow the temperature of the
  photosphere throughout the PRE phase, so any observed dip in the light curve
  is likely due to a dip in bolometric flux \citep{Keek2018b, Jaisawal2019}.  
  We investigate this in Section \ref{sec:time resolved}.

  Second, there is a noticeable pause during the rise to the first peak. Initially
  the flux increases rapidly; however, between $t\simeq0.6\s$ and $t\simeq1.3\s$, this
  rise briefly stalls, with the count-rate remaining constant at
  $\approx13,600$\,\cts.  After this pause, the rate continues to increase
  toward the maximum, albeit at a slightly slower pace (Figure \ref{fig:lc},
  inset of top panel). Simultaneously, the hardness ratio (the $3-10$\kev rate over
  the $0.3-1$\kev rate) evolves dramatically.  As the count-rate begins to
  rise, the hardness ratio spikes. Subsequently, the hardness briefly dips, and
  then stabilizes. It is during the dip that the pause in count-rate is
  observed. Additionally, the previously mentioned dip in count-rate (at
  $t\simeq13\s$), coincides with a similar dip in the hardness ratio,
  suggesting these two features are related.

\subsection{Pre-burst emission}
\label{sec:preburst}
  We extracted a spectrum from $400\s$ prior to the burst and modelled it in
  \textsc{xspec} v12.10 \citep{Arnaud1996}. Following \citet{DiSalvo2019}, we
  find that the spectrum could be well described with the model
  \begin{quote}
      \texttt{tbabs( diskbb + nthcomp )},
  \end{quote}
  where the \texttt{tbabs} interstellar absorption model \citep{Wilms2000} was
  used with the photoelectric cross-sections of \citet{Verner1996},
  \texttt{diskbb} \citep{Makishima1986} is a multi-color disk blackbody
  component, and \texttt{nthcomp} \citep{Zdziarski1996, Zycki1999} is a thermal
  Comptonization component.  We used an absorption column density of $N_H =
  2.1\E{21}$\,\persq{cm} \citep{Papitto2009, DiSalvo2019}, and electron
  temperature of {30}\,keV \citep{DiSalvo2019}. We further tied the
  \texttt{nthcomp} photon seed temperature to the disk temperature. Our
  best-fit had a reduced $\chi^2$ (\rchi) of 1.04 for 375 degrees of freedom
  (dof), yielding an inner disk temperature of $kT=0.70\pm0.07$ keV and a power-law
  photon index of $\Gamma=2.0\pm0.4$.  Using the \texttt{cflux} model we
  further measured the unabsorbed $1-10$ keV flux to be $(2.85 \pm
  0.05)\E{-10}\,\fluxcgs$, from which we extrapolate a bolometric flux
  ($0.01-100$\,keV) of {$(4.7 \pm 0.5)\E{-10}\,\fluxcgs$}.  Assuming a distance
  of $3.5$\,kpc \citep{Galloway2006}, a 1.4\,\msol neutron star mass, and a
  10\,km radius, this corresponds to a mass accretion rate of $\dot M =
  2.9\E{-11}\,\msol$\per{yr}, which is $\approx0.3\%$ of the Eddington rate.

\subsection{Burst spectroscopy}
\label{sec:spectra}

  \begin{figure}[t]
      \centering
      \includegraphics[width=\linewidth]{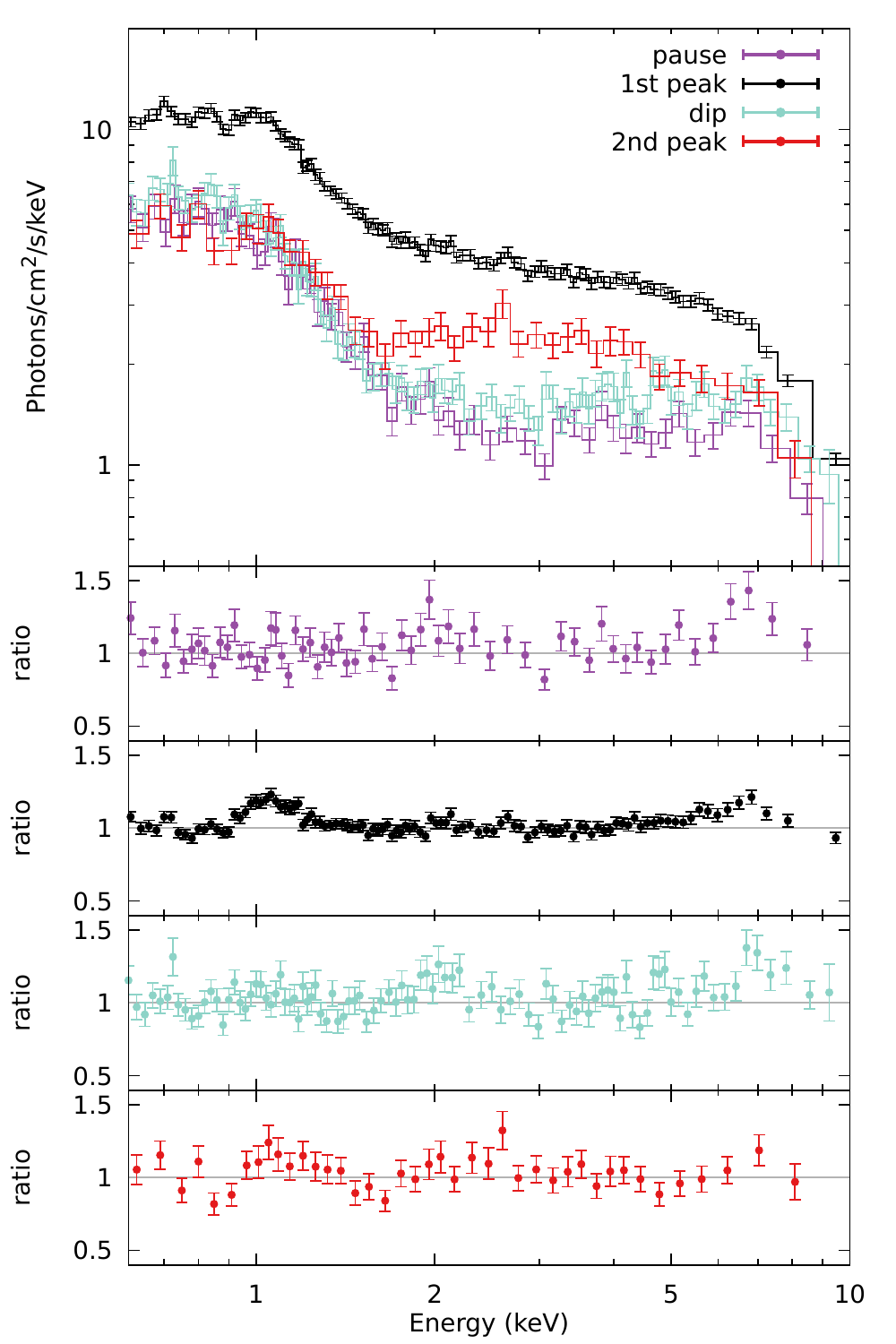}
      \caption{%
          Top: burst spectra at four epochs during the X-ray burst. Bottom: the
          residuals of the best fit models, showing the ratio of the data over
          the model prediction. In each case we have set the
          normalization of the emission line components to zero, which
          highlights these lines in the residuals.
      }
      \label{fig:spectra}
  \end{figure}

  We investigated the spectral shape of the burst emission by extracting a
  spectrum from a $4\s$ interval at the peak of the burst (Figure
  \ref{fig:spectra}).  We first attempted to model this spectrum by adding a
  blackbody component to the preburst spectrum model, holding all parameters
  except for those of the blackbody constant. At a $\rchi$ of $28$ for 631 dof,
  this model failed to account for a large excess below 1.5 keV and above 5
  keV.

  In an attempt to account for the residuals we applied a free scaling factor
  to the components describing the preburst spectrum \citep{Worpel2013}, so that 
  our model was
  \begin{quote}
      \texttt{tbabs(bbodyrad + f(diskbb + nthcomp))},
  \end{quote}
  where \texttt{bbodyrad} is a blackbody component with its normalization
  proportional to surface area. At a \rchi of 5.4 for 630 dof, this model
  failed to remove the large residuals. Additionally, at $f=159$, the magnitude
  of the obtained scaling factor is not realistic, as it is much larger than
  the ${f\sim2-10}$ that is typically observed \citep{Worpel2013}. 

  In an alternative approach to account for the large soft excess, we adopted a model
  consisting of the fixed preburst model plus two blackbody components. This model
  provided a much better description of the data (\rchi of 1.11 for 628), yielding
  a blackbody with $0.233\pm0.003$\,keV and $318\pm5$\,km for the
  soft excess, and $1.83\pm0.03$\,keV and $14.7\pm0.3$\,km for the higher
  energy emission (presumably the photosphere). 
  Some structure still remained in the residuals, most prominently at 1.0\,keV and 6.5\,keV.
  The fit was significantly improved (\rchi of 1.07 for 624 dof) by
  adding a \texttt{diskline} component \citep{Fabian1989} at
  $6.7_{-0.3}^{+0.1}$\,keV, along with a Gaussian line at $1.05\pm0.02$\,keV.
  The signal strength of the \texttt{diskline} was insufficient to reliably
  constrain the disk radius and inclination, giving respective limits of
  $<13\,r_g$ and $>65\arcdeg$. Instead, we fixed the inner radius to $11\,r_g$,
  which is the approximate magnetospheric radius \citep{Bult2015b}, and the
  inclination to $65\arcdeg$, which is within the range allowed by modelling
  of the Fe\,K line in the persistent emission of \src \citep{Cackett2009, Papitto2009, 
  DiSalvo2019}. 
  We further note that while this inclination is inconsistent with the
  $\leq30\arcdeg$ limit derived by \citet{Galloway2006}, a more sophisticated
  analysis of the same burst data yield $69_{-2}^{+4}\arcdeg$
  \citep{Goodwin2019}.  With these parameters held constant, we obtained a line
  normalization of $0.62\pm0.16$\phnorm. Meanwhile, the 1\kev Gaussian line
  had a normalization of $0.27\pm0.07$\phnorm and standard deviation of
  $0.05_{-0.02}^{+0.07}$\,keV.

  In an attempt to apply a physical foundation to our modelling of these data, we
  also fit the spectrum using the reflection models of \citet{Ballantyne2004b}.
  This overall model is summarized as
  \begin{quote}
      \texttt{tbabs( bbodyrad + diskbb + nthcomp \newline
      \phantom{tbabs}+ rdblur*atable\{reflection\} )}
  \end{quote}
  where \texttt{rdblur} is a convolution component that applies the
  relativistic effects associated with an accretion disk around a compact
  object, and the \texttt{reflection} component tabulates reflection spectra
  calculated for a hydrogen density of $n_H=10^{18}\,{\rm cm}^{-3}$, using a
  grid in temperature ($kT$), ionization ($\log \xi$), and Fe abundance ($\log
  {\rm Fe}$).  As before, we kept the parameters for the absorption column and
  preburst components fixed. The temperature parameter of the reflection table
  was linked to the blackbody temperature, and the \texttt{rdblur} parameters were
  identical to those of the \texttt{diskline} component discussed above. This
  model yielded a reasonable description of the continuum (\rchi of 1.2 for 628
  dof), but left a large residual at 1 keV. Adding in a Gaussian component gave
  a good fit to the data at a \rchi of 0.95 for 633 dof, with a normalization
  of $0.86\pm0.15$\phnorm and standard deviation
  $0.09\pm0.02$\kev. The best-fit parameters for the reflection component were
  $\log \xi=3.79_{-0.08}^{+0.11}$, $\log {\rm Fe}=0.51_{-0.24}^{+0.10}$, and an
  unabsorbed bolometric reflection flux of $(1.87\pm0.13)\E{-7}$\,\fluxcgs,
  indicating the reflection fraction is $F_{\rm refl} / F_{\rm bb} \approx2.3$.
  Finally, we note that in this model, the photosphere blackbody had a
  temperature of $2.05\pm0.06$\,keV and a radius of $6.6\pm0.7$\,km. While the
  temperature is consistent with the double blackbody model, the radius is
  significantly smaller.

  The double blackbody and reflection models both provide a statistically
  acceptable description of the spectrum, and give a roughly equivalent
  interpretation.  Since the double blackbody model is phenomenologically
  simpler, it can be fit robustly to much shorter integration times, yielding a
  higher time resolution view of the spectral evolution in the burst. In the
  following we will therefore focus our analysis on this model. 

  We extracted spectra from three other distinct time intervals during
  the burst: the pause (0.6\s), the dip (0.7\s), and the second peak (2\s).
  Each spectrum could be described with the double blackbody spectrum plus
  $6.7$\kev \texttt{diskline} component. The second peak additionally required
  an emission line at 1 keV. These spectra are shown in Figure \ref{fig:spectra}
  and their best-fit parameters are listed in Table \ref{tab:spectra}.

  \begin{table*}
   \newcommand{\mc}[1]{{#1}}
   \centering
   \caption{%
       Spectroscopy of the X-ray burst.
   \label{tab:spectra}
   }
   \newcommand{\ch}[1]{\colhead{#1}}
   \newcommand{\dch}[1]{\dcolhead{#1}}
   \newcommand{\ta}{\tablenotemark{a}}
   \newcommand{\tb}{\tablenotemark{b}}
   \begin{tabular}{ l l l C C C C }
    \decimals
    \tableline
    \ch{Component} & \ch{Parameter} & \ch{Unit} & \ch{Main Peak} & \ch{Pause} & \ch{Dip} & \ch{Second peak} \\
    \tableline
       Soft blackbody & temperature   & keV     & 0.233\pm0.003 & 0.228\pm0.007 & 0.210\pm0.005 & 0.22\pm0.01 \\
       Soft blackbody & normalization & km      & 318 \pm 5     & 240\pm17      & 297\pm18      & 236\pm31 \\[0.5em]

       Hard blackbody & temperature   & keV     & 1.83\pm0.03   & 2.5\pm0.2 & 2.52\pm0.14 & 1.99\pm0.11 \\
       Hard blackbody & normalization & km      & 14.7\pm0.3    & 6.2\pm0.5 & 6.8\pm0.4   & 10.5\pm0.7 \\[0.5em]

       Gaussian line  & line energy   & keV     & 1.05\pm0.02         &     &     & 1.05\pm0.02 \\
       Gaussian line  & standard deviation& keV & 0.05_{-0.02}^{+0.07}&     &     & 0.07\pm0.02 \\
       Gaussian line  & normalization & \phnorm & 0.27\pm0.07         &     &     & 0.26\pm0.07 \\[0.5em]

       diskline       & line energy   & keV     & 6.7_{-0.3}^{+0.1} & 6.7\pm0.1   & 6.7\pm0.1   & 6.8\pm0.1 \\
       diskline       & inclination   & degrees & 65                & 65          & 65          & 65  \\
       diskline       & inner radius  & GM/c$^2$& 11                & 11          & 11          & 11       \\
       diskline       & normalization & \phnorm & 0.62\pm0.16       & 0.47\pm0.14 & 0.5\pm0.2   & 0.45\pm0.16 \\
    \tableline
       \rchi / dof    &               &         & 1.07/624 & 1.13/182 & 1.05/264 & 1.17/487 \\
    \tableline
   \end{tabular}
   \flushleft
   \tablecomments{%
       Best-fit parameters of the spectral modelling described in Section
       \ref{sec:spectra}. Uncertainties are quoted at $90\%$ confidence. If no
       uncertainty is given, the parameter was held fixed. If no value is listed, 
       then the component was not included in the model.
   }
  \end{table*}

\subsection{Time-resolved spectroscopy}
\label{sec:time resolved}
\begin{figure}[t]
      \centering
      \includegraphics[width=\linewidth]{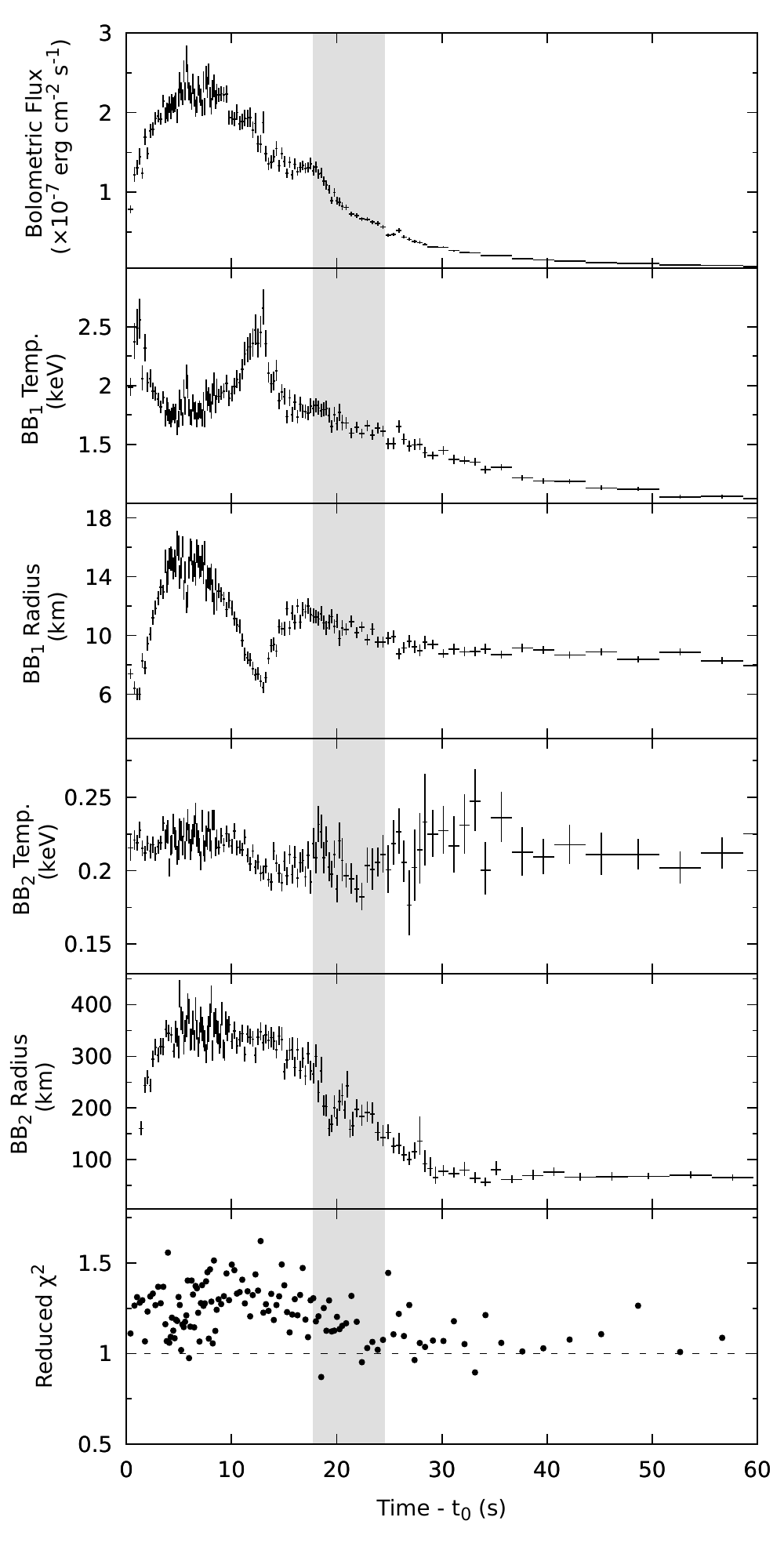}
      \caption{%
          Time-resolved spectroscopy of the X-ray burst using a double blackbody
          model. The top panel shows the estimated bolometric flux, the middle
          four panels show the time evolution of the spectral parameters
          describing the burst emission (radii were calculated using a distance
          of 3.5\,kpc), and the bottom panel gives the reduced $\chi^2$ fit
          statistic. The grey band indicates the time interval where burst
          oscillations were detected. See text for further details.
      }
      \label{fig:time resolved}
  \end{figure}

  To investigate the full spectral evolution of the X-ray burst, we
  applied high-resolution time-resolved spectroscopy. We extracted {133}
  spectra from dynamically allocated intervals. Each interval was constructed
  to have at least {0.125\s} of exposure, and was increased as needed to
  include a minimum of {2000} counts. Each spectrum was then fit using the
  double blackbody model.  For simplicity we did not include emission line
  components, giving in slightly poorer fit statistics. The resulting
  evolution in spectral parameters is shown in Figure \ref{fig:time resolved}. 

  The time resolved spectroscopy demonstrates that the hot ($\approx2$\,keV)
  blackbody in our model can be understood as the emission from a neutron
  star photosphere that undergoes PRE between $t\simeq1\s$ and $t\simeq13\s$. The radius
  expansion is moderate, reaching a maximum radius of $\approx15$\,km. The cool
  ($\approx0.2$\,keV) blackbody, on the other hand, maintains a stable
  temperature through the burst, with its emitting area closely tracking the
  evolution of the overall flux. This trend further supports the idea that the
  soft excess tracks an interaction between the burst emission and the neutron
  star environment, such as the disk reflection model discussed in Section
  \ref{sec:spectra}.

  Considering the bolometric flux, we see that the burst emission reaches a stable peak of
  $(2.40\pm0.12)\E{-7}$\,\fluxcgs when the photosphere is at its largest
  extent. As the photosphere begins to contract, the flux begins to decrease.
  This cooling trend, however, is interrupted at $t\simeq15.5$. In the following
  $\approx3\s$, we see the bolometric flux holding constant, causing
  an excess over the cooling trend that coincides exactly with the second
  peak observed in the light curve. Hence, the spectroscopy confirms that the
  dip and second peak seen in the light curve are indeed astrophysical in
  origin rather than a passband effect.
  
  Finally, we note a peculiar feature in the spectroscopic results: the
  temperature evolution of the photosphere shows two peaks, marking the start and
  end of the PRE phase. These start and end times coincide with the pause and
  the dip, respectively. Furthermore, both the pause and the dip have the same bolometric 
  flux level of $(1.43\pm0.09)\E{-7}$\,\fluxcgs.

\subsection{Burst oscillations}
\label{sec:oscillations}
  To search for burst oscillations, we constructed a $1/8192\s$
  time-resolution light curve in the $0.3-10$ keV energy band. We then
  searched for coherent oscillations in a {10}\,Hz region centered on
  the known 401\,Hz pulsar spin frequency by applying a sliding window search
  method \citep[see, e.g.,][and references therein]{Bilous2019}. Specifically,
  we used window sizes of $T=1,2,{\rm~and~} 4\s$, with strides of
  $T/10\s$. For each window position, we computed a power spectrum and searched
  for a signal power in excess of the expected noise distribution
  \citep{Klis1989}.  We detected a significant oscillation ($>3\sigma$) in all
  windows between $t=17.7\s$ and $t=24.6\s$. The fractional sinusoidal
  amplitude\footnote{Sinusoidal amplitudes are a factor of $\sqrt{2}$ larger than fractional
  rms amplitudes.} of the burst oscillation was $(4.0\pm0.6)\%$, while the
  oscillation frequency was $401$\,Hz and did not show any significant drift.

\begin{figure}[t]
      \centering
      \includegraphics[width=\linewidth]{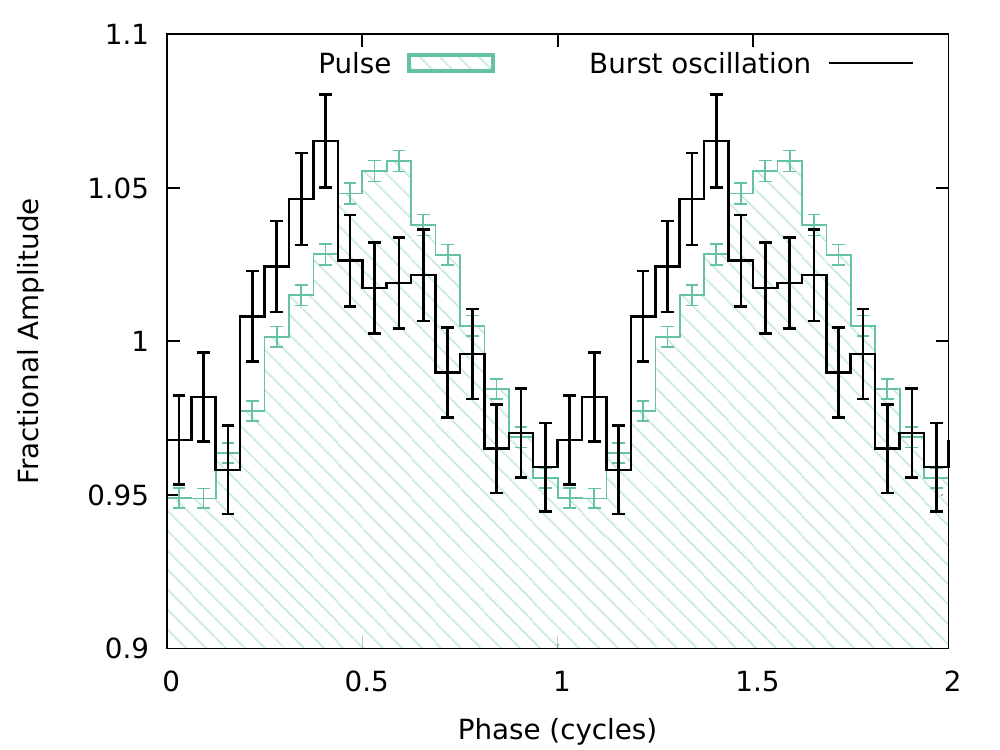}
      \caption{%
          Waveform of the burst oscillation observed in \src, compared
          to the waveform of the accretion-powered pulsation as seen outside
          the X-ray burst interval.
      }
      \label{fig:bo}
  \end{figure}

  Given the stability of the burst oscillation frequency, we folded the event times
  within the noted epoch on the pulsar spin period to obtain a waveform for the
  burst oscillation. For comparison, we also extracted the waveform of the
  persistent pulsations from the full ObsID, excluding the burst emission (see
  \citealt{ATelBult19c} for a preliminary ephemeris). Both waveforms are shown in
  Figure \ref{fig:bo}. The burst oscillation has a similar profile and
  amplitude as the persistent pulsation, but appears to lead the pulse by 
  $34\arcdeg\pm7\arcdeg$.

\begin{figure}[t]
      \centering
      \includegraphics[width=\linewidth]{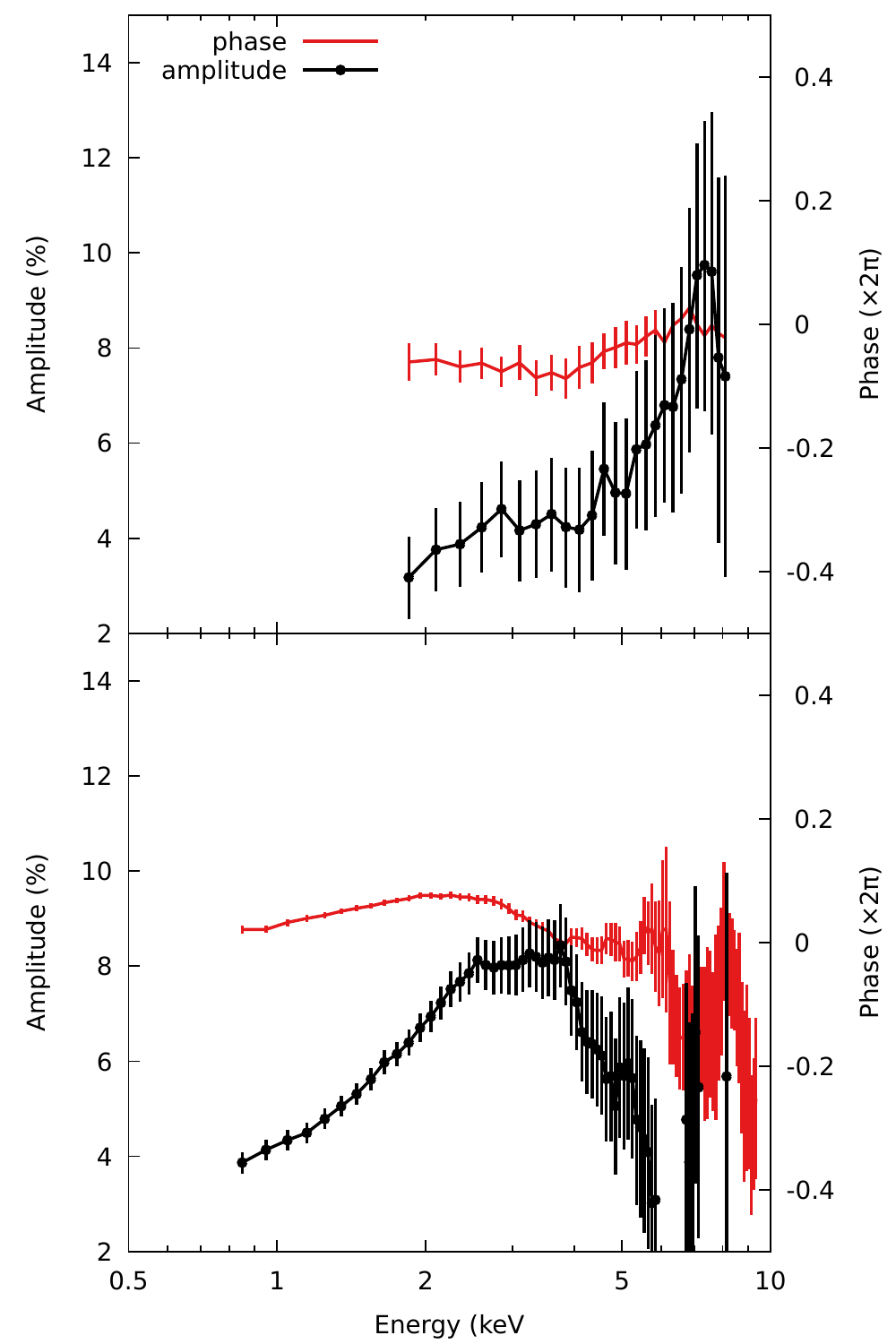}
      \caption{%
          Top: energy dependence of the burst oscillation waveform. Bottom: same for
          the persistent pulsation observed outside the X-ray burst interval. Note
          that these data were computed using a sliding window, so adjacent points are
          correlated. Additionally, in order to look for decohering noise, we
          plot the waveform phase even if the amplitude is not formally
          significant. See text for further details.
      }
      \label{fig:bo-spectrum}
  \end{figure}

  {
  To resolve the burst oscillation with respect to photon energy, we applied a
  sliding window to the instrument energy channel space, using a window size of
  100 channels and strides of 10 channels. At each window position, we folded
  the selected data and measured the burst oscillation amplitude and phase. We
  then repeated this method for the non-burst data. The resulting
  amplitude spectra are shown in Figure \ref{fig:bo-spectrum}. Although the averaged
  profiles are similar, the energy dependence of the burst oscillation is very
  different from that of the pulsar. Particularly notable is that the burst
  oscillation amplitude is mostly constant below $\approx5$\,keV, but rises
  sharply at $6-7$\,keV.
  }

\section{Discussion}
\label{sec:discussion}
  We detected a bright X-ray burst from \src with \nicer. The burst showed a 
  peculiar light curve, with a notable pause during the rise and a double-peaked
  structure. Additionally, we detected significant burst oscillations in the
  cooling tail of the burst and emission features in the burst spectrum. We now
  discuss each of these findings.
  
  \subsection{Reflection}
    We find that the burst spectrum shows a strong excess at the lowest
    energies that requires the inclusion of a second blackbody in the spectral
    model. A similar strong soft excess was previously observed in \src by
    \citet{Zand2013}. In contrast to that work, however, we also observe 
    emission features at 1\, keV and 6.7\,keV. A similar complex of 
    emission features has been seen in the intermediate-duration X-ray
    bursts from IGR~J17062--6143 \citep{Degenaar2013,Keek2017}, which were
    associated with ionized Fe\,L and Fe\,K emission lines. Thus,
    the detection of these lines provides strong evidence
    that we are seeing the burst emission reflected from the accretion
    disk.  Applying a physically motivated disk reflection model to our data
    (Section \ref{sec:spectra}) indicates that such a reflection
    component provides a satisfactory description of the soft excess, but cannot
    fully account for the emission feature observed at 1\,keV. We suggest that
    this may be due to the presence of additional elements not currently
    incorporated in these models (e.g. Ne). Additionally, this model fit
    indicated a strong reflection signal, which may indicate that the accretion
    disk structure is significantly impacted by the burst \citep{He2016,
    Fragile2018}.

  \subsection{Spectral evolution}
    The X-ray burst light curve shows a double-peaked structure. Given that the
    second peak in count-rate occurs after the PRE phase, and that this peak is
    reproduced (albeit less prominently) in the bolometric flux, we conclude
    that this feature is astrophysical in origin. A very similar double-peaked
    structure in an X-ray burst from 4U 1608--52 was recently observed with
    \nicer \citep{Jaisawal2019}.  Although that burst showed a hotter
    photosphere and lacked the strong soft excess that we detect in \src, the
    rebrightening phase is nearly identical in both bursts. In each case, the
    end of the PRE phase coincides with a pronounced dip in count-rate, and is
    followed by a secondary peak. 
    
    \citet{Jaisawal2019} considered a number of plausible origins for the rebrightening
    of the burst flux, including the ignition of fresh material \citep{Keek2017b}, stalled
    thermonuclear flame spreading \citep{Bhattacharyya2006}, and waiting points in the
    rp-process \citep{Fisker2004}. Our observation of rebrightening in the X-ray burst
    from \src does not rule out any of these proposed explanations. It does, however,
    add two new perspectives: first, in \src the rebrightening coincides with the onset
    of burst oscillations, which may be difficult to reconcile a flame
    spreading model. Second, in \src the dip appears related to the pause during the rise. If 
    this relation is real, then whatever physical mechanism underpins these
    features may also be related to the rebrightening.

  \subsection{Burst Oscillations}
    We found that the X-ray burst shows burst oscillations at the expected 401 Hz
    spin frequency. Comparing these oscillations with the accretion-powered pulsations, we
    find that the two waveforms are remarkably similar, but the burst oscillations
    lead the pulsations by {$34\arcdeg\pm7\arcdeg$}. Similar results were reported
    from \rxte observations of burst oscillations in \src observed during the cooling
    phase of an X-ray burst \citep{Chakrabarty2003}. The fact that the burst oscillations
    are so closely matched to the persistent pulsations in terms of their waveform, suggests
    that both must arise from geometrically similar, if not the same, confined emitting
    region (hot-spot) on the stellar surface. With this in mind, it is interesting to note
    that the \nicer data indicate the waveform energy dependence of the burst oscillations is 
    quite different from that of the persistent pulsations. 
    Some of the difference may simply arise from the strong reflection component,
    which is likely not pulsed, and thus is expected to dilute the measured
    burst oscillation amplitude at low energies. A detailed spectral-timing
    analysis may be able to determine how much each of the spectral components is
    oscillating. Such an analysis, however, is beyond the scope of this initial
    work.

  \subsection{Eddington limits}
   Finally, we note that our analysis of the light curve, the spectral
   evolution, and the timing behavior all indicate that each time interval
   where the \nicer count-rate of \src is between 13,000 \cts and 14,000 \cts is
   somehow special. At these count rates, the burst rise pauses, the dip reaches
   its minimum, and burst oscillations appear.  The bolometric flux measured at
   these times was $(1.43\pm0.09)\E{-7}$\,\fluxcgs, which corresponds to
   a luminosity of $(2.08\pm0.13)\E{38}$\,\lumcgs.
   We note that this luminosity is consistent with the expected local
   Eddington limit of a hydrogen envelope of a neutron star \citep{Lewin1993}. 

  For spherically symmetric emission, the Eddington luminosity as measured by
  the observer is predicted as \citep{Lewin1993, Suleimanov2017}
  \begin{equation}
      L_{\rm edd} = \frac{4\pi GMc}{\kappa_{\rm T}}\frac{1}{(1+z)},
  \end{equation}
  where $M$ is the neutron star mass, $G$ the gravitational constant,
  $c$ the speed of light, and $z$ the gravitational redshift. Finally,
  $\kappa_{\rm T}=0.2(1+X)\,{\rm cm}^2$\per{g} is the Thomson electron scattering
  opacity, with $X$ the hydrogen abundance in the atmospheric layer. 
  Hence, we expect that the flux level at which PRE occurs differs depending on
  the material composition of the neutron star envelope.  For a hydrogen layer
  with cosmic abundances ($X=0.73$), the predicted luminosity is
  $\approx2.0\E{38}$\,\lumcgs, whereas the
  luminosity of a pure helium layer ($X=0$) is larger by a factor of 1.73. In
  \src, the ratio in bolometric flux between the peak and pause/dip is
  $1.68\pm0.13$.

  The following scenario now emerges for the X-ray burst evolution. As the
  critical ignition point is reached in the helium layer, the flame front
  quickly spreads across the stellar surface and an intense radiation field
  starts to diffuse outward. After about 0.5\s, the radiation pressure
  reaches the local Eddington limit of the hydrogen layer, causing that layer to
  expand. Meanwhile, the intensity of the radiation field continues to increase,
  either expelling or diluting the hydrogen layer, so that the observed
  spectrum becomes dominated the PRE of the helium layer. Over the following
  10\s, we observe the full helium PRE cycle, causing the photosphere to cool
  and then heat, as the envelope expands and then contracts. Once the envelope
  touches back down on the stellar surface, the burst flux is still comparable
  to hydrogen Eddington limit, although at the time this touchdown occurs,
  there will likely have been some mixing of the atmospheric layers.
  Subsequently, the photosphere cools, while the radius grows, and then, after
  about $4-5\s$, the rebrightening mechanism activates and quickly thereafter the
  burst oscillations appear. The photosphere continues to cool, and after about
  10\s both the rebrightening and oscillations switch off. 

  The flux levels of the pause and peak are highly suggestive that we are seeing
  both the hydrogen and helium Eddington limits in a single X-ray burst. The evolution
  of the hardness ratio around the pause further supports the interpretation that this
  stall in the rise is related to the expanding hydrogen layer. The link between
  the pause and the dip is weaker, but highly suggestive, and may yet provide the
  insight required to uncover the physics behind intrinsic rebrightening during
  the tail of an X-ray burst.

\nolinenumbers
\facilities{ADS, HEASARC, NICER}
\software{heasoft (v6.26), nicerdas (v6a), xspec (v12.10)}

\acknowledgments
This work was supported by NASA through the \nicer mission and the
Astrophysics Explorers Program, and made use of data and software 
provided by the High Energy Astrophysics Science Archive Research Center 
(HEASARC).
P.B. was supported by an NPP fellowship at NASA Goddard Space Flight Center. 
D.A. acknowledges support from the Royal Society. 

\bibliographystyle{fancyapj}

\end{document}